\documentclass[a4paper,reqno,12pt]{revtex4}

\usepackage{graphicx}
\usepackage{dcolumn}
\usepackage{epsfig}

\usepackage[centertags]{amsmath}
\usepackage{amsfonts}
\usepackage{euscript}
\usepackage{amssymb}
\usepackage{amsthm}
\usepackage{newlfont}
\usepackage{stmaryrd}
\usepackage{mathrsfs}
\usepackage{subfigure}

\newcommand{\bbN}{{\mathbb N}}

\newcommand{\bbZ}{{\mathbb Z}}

\newcommand{\opunit}{\text{1}\kern-0.22em\text{l}}

\DeclareMathAlphabet{\mathpzc}{OT1}{pzc}{m}{it}

\begin{document}

\title{Nonequilibrium physics aspects of\\
 probabilistic cellular automata}
\author{Christian Maes}
\email{christian.maes@fys.kuleuven.be}
\affiliation{Instituut voor Theoretische Fysica, KU Leuven, Belgium}
\maketitle
{\bf Probabilistic cellular automata (PCA) are used to model a variety of discrete spatially extended systems undergoing parallel-updating. We propose an embedding %They do not come easily however with a thermodynamic interpretation, perhaps as some claim, because they are most natural to consider as fundamental physics only on the Planck scale.  Nevertheless we attempt here to embed 
of a number of classical nonequilibrium concepts in the PCA-world.  We start from time-symmetric PCA, satisfying detailed balance, and we give their Kubo formula for linear response.  Close-to-detailed balance we investigate the form of the McLennan distribution and the minimum entropy production principle.  More generally, when time-symmetry is broken in the stationary process, there is a fluctuation symmetry for a corresponding entropy flux. For linear response around nonequilibria we also give the appropriate formula which is now not only entropic in nature.}

\section{PCA and physics}
Despite numerous programmes, ambitions and studies there is no derivation of the dynamics of probabilistic cellular automata (PCA) from more microscopic physical rules or from more fundamental physics as generally understood.  There is of course always the possibility to look at discrete time steps for a sequential-(continuous)-time interacting particle system, but that will not yield PCA as the latter are always {\it non-strategic} in the sense that their conditional probability given the previous configuration is a product distribution.   Moreover, from the point of view of continuous time, the discrete time-step appears to introduce another important time-scale into the physical problem, which would need to be accounted for. Alternatively there is of course always the possibility, repeatedly entertained,  that it is PCA that are {\it more} fundamental, and that the logic should in fact be reversed: the more standard physical descriptions must then be derived from PCA rules, \cite{nks}. In particular, thinking of PCA as physics on the Planck scale, classical space-time would emerge as a coarse-grained feature of quantum gravity, \cite{thooft}.  Whatever point of view, it is not automatic to transfer continuous time physical notions to the domain of PCA.  What has been done in the past is to connect $d$-dimensional PCA  with $(d+1)$-dimensional equilibrium statistical mechanical models, and we will review the main relation in the next section. For example, the study of phase transitions in PCA which may be useful for the understanding of robustness of large parallel computations, will benefit from (e.g. renormalization group) techniques in equilibrium statistical mechanics of critical phenomena.  The main motivation of the present paper is however to search for analogues of {\it non}equilibrium concepts, and to give PCA-versions of some recent results in stochastic kinetics.\\

Recently indeed much discussion was devoted to the application of a thermodynamic formalism to smooth dynamics \cite{rue}, including a thermodynamic discussion of stochastic processes modeling systems in weak contact with different equilibrium reservoirs, \cite{sekimoto,heatconduction}.  Nonequilibrium statistical physics is obviously expected to be incomplete when restricting it to such concepts as energy, work, heat and entropy(production) even when including the study of their fluctuations, but it is a good start to see how already these notions appear and play in physically motivated stochastic dynamics for open systems. The situation for PCA is then even worse.   Our basic method will not to start from detailed balance as expressed in terms of an energy function or potential, but rather begins with estimating time--reversal breaking.  That is the content of Sections \ref{db}--\ref{bdb}, where we repeat the fluctuation symmetry for the source term of time--symmetry breaking.  We then continue with that source term ``entropy production'' in the following sections where we discuss the minimum entropy production principle and the McLennan--Zubarev distribution.  We end by giving the linear response formula for general PCA.

\section{Notation}
We only consider translation-invariant probabilistic cellular automata on the cubic lattice $\bbZ^d$, characterized by the one site
updating
\[
p_i(a|\eta) = \mbox{Prob}[X_n(i) = a|X_{n-1}=\eta]
\]
for state space $K =\{+1,-1\}^{\bbZ^d}$, $a=\pm 1, \eta\in K$.  We refer to \cite{genr1,genr2} as general references.  $(X_n, n=0,1,2,\ldots)$ is  a discrete time Markov process on $K$, with
\[
\mbox{Prob}[X_n(i) = a_i, i\in V|X_{n-1}=\eta] = \prod_{i\in V}p_i(a_i|\eta) 
\]
for all finite $V\subset \bbZ^d$.  We prefer of course to have $p_i(a_i|\eta)$ to depend locally on neighboring $\eta(j), j\sim i$ only.

As a parameterization we choose to write
\begin{equation}\label{par}
p_i(a|\eta) = \frac 1{2}(1 + a\,h_i(\eta))
\end{equation}
where $|h_i|\leq 1$ on $d$-dimensional configurations.  Again, $h_i(\eta)$ is a local and translation invariant function of $\eta\in K$.
The formal $(d+1)$-dimensional Hamiltonian is
\[
H(\sigma) = -\sum_{i,n}\log p_{i,n}(\sigma_n(i)|\sigma_{n-1})
\]
for $\sigma = (\sigma_n(i), i\in \bbZ^d, n\in \bbN)$.  For local PCA the relative Hamiltonian $H(\sigma) - H(\sigma')$ where $\sigma=\sigma'$ outside some finite volume $\Lambda\subset \bbZ^{d+1}$ makes mathematical sense. That is in fact the start of the connection between $(d+1)$-dimensional equilibrium statistical mechanics and PCA as dynamics on discrete configurations on $\bbZ^d$, \cite{fromPCA,led}.  The present paper will emphasize the \emph{non}equilibrium aspects, and these start from realizing that the Hamiltonian $H$ does not need to be reflection--invariant in the temporal (or, $(d+1)$th)--direction.

\section{Detailed balance}\label{db}
In contrast to continuous time interacting particle systems, PCA as defined above cannot produce any given Gibbs distribution as stationary. In particular, detailed balance is not so naturally installed for PCA.  Remember indeed that the updating is in parallel with each spin being updated independently given the previous configuration, so that is it is not immediate how to minimize an energy function, or how to install a Lyapunov function, especially with local interactions. The change of $X_n(i)$ can be determined by some cost function ${\cal L}(X_n(i), X_n(j), j\sim i)$ but while $X_n(i)\rightarrow X_{n+1}(i)$ changes also its neighbors $X_n(j)\rightarrow X_{n+1}(j)$ get updated similarly and simultaneously, which may prevent gradient flow.   That is not to say that we cannot build invertible cellular automata, indeed we can, \cite{toffoli}, but the very concept of (semi-bounded) energy appears deeply related to a continuous time process.  Constructions involving the Hamiltonian formalism for integer-valued variables and integer time steps, are, to say the least, quite cumbersome.\\
Coming back to {\it probabilistic} cellular automata, a stationary process $(X_n, n\in \bbZ)$, is time-reversible (statistically symmetric under $n\rightarrow -n$) when in \eqref{par}
\[
h_i(\eta) = \tanh[\lambda_i +\sum_j J_{ij}\eta_j]
\]
for some $\lambda_i$ and symmetric $J_{ij}=J_{ji}$.
The stationary distribution is then, formally,
\[
\nu(\eta) = C\exp\sum_i\{\lambda_i\eta_i + \log 2\cosh[\lambda_i + \sum_j J_{ij}\eta_j]\}
\]
Note in fact that the corresponding interaction has at least  three--body interaction; to obtain a simpler nearest neighbor-interaction appears impossible.  It is then also true that, in contrast with continuous (sequential) time, not all equilibrium distribution can be reached as stationary distribution. For example, the standard Ising model cannot be obtained; see however \cite{seeh,scop}. An alternative is working on bipartite lattices, with alternate updating in the way of \cite{Domany1984}.\\

Detailed balance can formally be written as the condition that, pretending first we have a Markov chain with transition probability $p(\eta|\eta')$,
\[
p(\eta|\eta') \nu(\eta') = S(\eta,\eta')\]
is symmetric.  As a consequence then,
\[
\nu(\eta) = \nu(-1)\,\prod_i \frac{1+\eta_i h_i(-1)}{1-h_i(\eta)}
\]
where ``$-1$'' stands for the configuration which is constant equal to $-1$, see \cite{stavs73}.
That gives rise to a well--defined Hamiltonian on $K$.  We call such $\nu$ equilibrium distributions even though there is no thermodynamic notion of equilibrium here.  We can for example examine what happens to them under a small perturbation.
We are then talking about linear response around equilibrium.\\
Suppose we start in equilibrium (with expectations $\langle\cdot\rangle_{\text{eq}}$) and we perturb  ($\rightarrow \langle \cdot\rangle_{\text{eq}}^h$) by letting
\begin{equation}\label{pert}
p^h_i(\sigma_n(i)|\sigma_{n-1}) = \frac{p_i(\sigma_n(i)|\sigma_{n-1})}{z_i(\sigma_{n-1})} \,e^{\frac{h_n}{2}[V_i(\sigma_n)-V_i(\sigma_{n-1})]}\quad n=1,2\ldots
\end{equation}
where all $V_i$ are local and only a finite number are non-zero, and the $h_n$ are small amplitudes.  The linear response 
on an observable $O$ at time $n>m$ is found to be
\begin{equation}\label{ku}
\frac{\partial}{\partial h_m}\langle O(\sigma_n)\rangle^h_{\text{eq}}\,(h=0) = \frac 1{2} \sum_i \langle[V_i(\sigma_{m+1})- V_i(\sigma_{m-1})]\,O(\sigma_n)\rangle_{\text{eq}}
\end{equation}
where the subscript reminds us that the reference (unperturbed) process is equilibrium time-reversal symmetric.  The right--hand side is an equilibrium time--correlation function.
We recognize the analogue of the Kubo formula (or the fluctuation--dissipation theorem) around equilibrium, \cite{kubo}.

\section{Breaking detailed balance}\label{bdb}
A measure for breaking detailed balance is given by
\begin{equation}\label{je}
J_{i,n}(\sigma) := \log \frac{ p_i(\sigma_n(i)|\sigma_{n-1}) }{ p_i(\sigma_{n-1}(i)|\sigma_n) }
\end{equation}
which is a local function on $\bbZ^{d+1}$ (involving just two-time layers).  The reason is that there is always 
$G_{L,N}(\sigma)$ with uniform bound $||G_{L,N}||\leq c(d)NL^{d-1}$ so that
\[
W_{N,L}(\sigma):=\sum_{n=-N+1}^{N-1}\sum_{|i|\leq L-1} J_{i,n}(\sigma) + G_{L,N}(\sigma)
\]
is antisymmetric under time--reversal $(\theta_{L,N}\sigma)_n(i) := \sigma_{-n}(i)$ for $(i,n) \in \Lambda_{L,N}$ which is a rectangular shaped region centered
at the origin with time-extension $2N+1$ and spatial volume $(2L+1)^d$. Under detailed balance, for the equilibrium process then $\langle W_{N,L}\rangle_{\text{eq}} =0$.\\

There is actually a further symmetry, called fluctuation symmetry, in the following sense:\\
For $L=L(N)\leq N$ growing to infinity with time $N$, the limit
\begin{equation}\label{el}
e(\lambda) := \lim_N \frac 1{|\Lambda_{L,N}|}\log\langle e^{-\lambda \sum_{(i,n)\in \Lambda_{L-1,N-1}} J_{i,n}}\rangle
\end{equation}
exists for all real $\lambda$ and $e(\lambda)=e(1-\lambda)$.  The expectation $\langle\cdot\rangle$ is for a general local PCA in the stationary regime.  We refer to \cite{gibbsproperty,jmp2000} for a proof and extensions within the context of Gallavotti--Cohen symmetries, \cite{gc}.

\section{Entropy production rate density}

For a stationary distribution $\nu$ we consider its extension (the stationary Markov process) $P_\nu$ on $\bbZ^{d+1}$.  In analogy with continuous time \cite{jmp2000}, we \emph{define} the mean entropy production rate per unit volume as the space--time relative entropy density with respect to time-reversal
\[
\mbox{MEP}_\nu := s(P_\nu|P_\nu\theta)= - s(P_\nu) + \langle \log p_0(\sigma_0(0)|\sigma_1) \rangle = \langle J_0\rangle
\]
where $J_0$ is found from \eqref{je} with $i=0=n$ and $s(P_\nu)$ is the statistical mechanical equilibrium entropy of the $(d+1)$dimensional Gibbs measure, also called Kolmogorov-Sinai entropy,
\[
s(P_\nu)= -\langle \sum_a p_0(a|\sigma_{-1})\log p_0(a|\sigma_{-1})\rangle
\]
Whether MEP$_\nu$ truly corresponds to an entropy production is unclear, as we have not obtained PCA as subsystem or as reduced description after contact with heat baths etc.  It is rather to be seen here as the expected rate of time-reversal breaking.  Clearly MEP$_\nu$ is non-negative, and equals zero at detailed balance.
It is the first $\lambda$-derivative of $e(\lambda)$ in \eqref{el}.\\

If the process is not stationary but has reached probability distribution $\mu$, we  define the expected entropy production rate in $\mu$ as
\[
\mbox{EP}[\mu]:= \langle J_0(\sigma)\rangle_{\mu} + S(\mu P) - S(\mu)
\]
where the first term takes the expectation of \eqref{je} over the two-time layer $(\sigma_0,\sigma_1)$ when $\sigma_0$ is averaged with probability distribution $\mu$ on $K$.  The $S(\mu)$ and the $S(\mu P)$ are
Shannon entropy densities for $\mu$ and its (single step)  update $\mu P$ (with stochastic matrix $P$).  In fact we can also write EP$[\mu]$ itself as a relative entropy density of $P_\mu$ restricted to two-time layers with respect to
$P_{\mu P} \theta$ on these two times and with $\theta$ exchanging the two times, or formally
\begin{equation}\label{epf}
\mbox{EP}[\mu] \propto \sum_{\sigma_0,\sigma_1}\mu(\sigma_0)p(\sigma_1|\sigma_0)\log\frac{\mu(\sigma_0)p(\sigma_1|\sigma_0)}{\mu P(\sigma_1)p(\sigma_0|\sigma_1)}
\end{equation}
 The functional EP$[\mu]$ is non-negative, convex and vanishes under detailed balance when $\mu$ is the equilibrium distribution.  There is in fact a unique minimizer, which we could call the Prigogine distribution.

\section{Minimum entropy production principle}
It turns out that  when operating close to detailed balance the stationary distribution can also be characterized as minimum of a functional which very much resembles the entropy production rate density.  In other words the stationary $\nu$ equals a minimizer of an entropy production-like functional.\\
We give here the argument for any fixed finite volume (perhaps with periodic boundary conditions) on which the PCA gets defined, which is the case of a (discrete time) Markov chain.  We follow below the straightforward variational method of \cite{vdb}.
Whether the minimum entropy production principle or a close relative of it can also be derived as a consequence of dynamical large deviation theory in the way of \cite{minep}, remains an open question.\\

Consider
\[
\sigma[\mu] := \sum_{x,y}\mu(x)p(x,y)\log\frac{\mu(x)p(x,y)}{\mu(y)p(y,x)}
\]
and take the variation with respect to $\mu(x)$ to find
\begin{equation}\label{var}
\sum_{y}p(x,y)\log\frac{\mu(x)p(x,y)}{\mu(y)p(y,x)}  -  \frac{\mu P(x)}{\mu(x)} = \text{constant} 
\end{equation}
We now like to show that \eqref{var} is indeed satisfied to first order around equilibrium. The latter is quantified via a dimensionless parameter $\varepsilon\ll 1$. We take $\mu = \nu(1 +\varepsilon g)$ and 
$p(x,y) = t(x,y)(1 + \varepsilon m(x,y))$ with detailed balance for $\nu(x) t(x,y) = t(y,x) \nu(y)$.   Then the first term in the left-hand side of \eqref{var} becomes
\[
\sum_y t(x,y)(1 + \varepsilon m(x,y)) \log\frac{(1+ \varepsilon g(x))(1+ \varepsilon m(x,y))}{(1+\varepsilon g(y))(1+ \varepsilon m(y,x))} =1 +\varepsilon v(x)
\]
(expanding to first order in $\varepsilon$) where
\begin{equation}\label{ve}
v(x) =  \sum_y t(x,y)[g(x)+ m(x,y) - g(y) - m(y,x)] = g(x) - \sum_y t(x,y) g(y) - \sum_y t(x,y) m(y,x)
\end{equation}
from using $\sum_y t(x,y) = 1$ and $\sum_y t(x,y) m(x,y) = 0$.  
The second term in \eqref{var} contains
$\mu P(x) = \nu(x)(1+ \varepsilon\tilde g(x))$, where
\begin{eqnarray*}
 \nu(x)(1+ \varepsilon\tilde g(x)) &=& \sum_y t(y,x)(1 + \varepsilon m(y,x)) \nu(y) (1 +\varepsilon g(y))\nonumber\\
  &=& \nu(x) + \varepsilon\sum_y t(y,x)  m(y,x) \nu(y) + \varepsilon\sum_y t(y,x) \nu(y) g(y)\nonumber\\
 &=&  \nu(x) + \varepsilon \nu(x) \sum_y t(x,y)  m(y,x) + \varepsilon\nu(x)\sum_y t(x,y) g(y)\nonumber\\
 &\implies& \tilde g(x) =  \sum_y t(x,y)  [m(y,x) +  g(y)]
 \end{eqnarray*}
where we used detailed balance $t(y,x)\nu(y) = \nu(x) t(x,y)$. Therefore,
\[
\frac{\mu P(x)}{\mu(x)} = 1+ \varepsilon\tilde g(x) - \varepsilon g(x) = 1 +\varepsilon [\sum_y t(x,y)  [m(y,x) +  g(y)] - g(x)]
\]
which we must compare with \eqref{ve} to see that indeed \eqref{var} is satisfied.\\

Remark that $\sigma[\mu]$ not quite equals \eqref{epf}  for $\sigma_0\rightarrow x, \sigma_1\rightarrow y, p(\sigma_1|\sigma_0)\rightarrow p(x,y)$.
We really would have to consider instead of $\sigma[\mu]$ the entropy production functional
\[
\mbox{EP}[\mu] = \sum_{x,y}\mu(x)p(x,y)\log\frac{\mu(x)p(x,y)}{\mu P(y)p(y,x)}
\]
However, taking the variation of that one, we find that the stationary distribution does \emph{not} satisfy it even to first order around equilibrium.  In other words we should not expect that the stationary distribution of a PCA equals the Prigogine distribution even in linear order.

\section{McLennan-Zubarev formula}
Close-to-detailed balance we can give an expression for the stationary distribution.  In \cite{mcL} is explained a rigorous derivation for continuous time.  Let us here look at a (discrete time, irreducible and aperiodic) Markov chain $X_n, n\geq 0$, for $X_n\in K$ finite.\\
From the previous section we know that the distribution $\mu$ coincides with the stationary distribution $\nu$ to linear order in $\varepsilon$ when it satisfies \eqref{var}.  So we can get $\mu$ correct to first order by plugging it in \eqref{var}:  (using $\mu=\mu P$),
\[
 \sum_{y}[p(x,y) - \delta_{x,y}]\log \mu(y) =  \sum_y p(x,y)\log \frac{p(x,y)}{p(y,x)} + \text{constant} 
\]
We substitute again $\mu(x) = \nu(x) (1+ \varepsilon g(x))$ and $p(x,y) = t(x,y)(1+ \varepsilon m(x,y))$ and we must have that \eqref{ve} is constant:
\begin{equation}\label{vel}
g(x) - \sum_y t(x,y) g(y) = \sum_y t(x,y) m(y,x) + \text{constant}
\end{equation}
which we must solve for $g$.  One should however be aware that the (detailed balance) matrix $L$ with element $t(x,y)-\delta_{x,y}$ is singular. We can however use the constant to project on the subspace orthogonal to the constant functions.  That is the so called pseudo-inverse $L^{-1}$ for which we have
\[
L^{-1}f(x) = -\sum_{n=0}^\infty P^n [f -\langle f\rangle_{\text{eq}}]
\] 
to be used for the function $f(x) = \sum_y t(x,y) m(y,x)$.  That gives the correction $g$ to equilibrium, yielding the McLennan--Zubarev form.  To work out the analogous McLennan-Zubarev form for PCA (in the thermodynamic limit) and to show it is a Gibbsian distribution at least in the high noise regime is left here as an open problem.

\section{Linear response}
For the perturbation \eqref{pert}, but now starting from a general distribution $\rho$ and not restricting ourselves to detailed balance, we have the nonequilibrium response formula
\begin{eqnarray}
\langle O(\sigma_n)\rangle^h_\rho - \langle O(\sigma_n)\rangle_\rho &=&
\sum_i\sum_{m=1}^{n-1}\frac{h_m}{2}\{\langle [V_i(\sigma_{m+1})- V_i(\sigma_m)]\,O(\sigma_n)\rangle_\rho \nonumber\\
&-&
\langle \langle V_i(\sigma_{m+1})- V_i(\sigma_m)|\sigma_m\rangle \,O(\sigma_n)\rangle_\rho\} + O(h^2)
\end{eqnarray}
It is the generalization of the Kubo-like formula \eqref{ku} to nonequilibrium processes.  It contains a frenetic contribution following the line of \cite{fdr}.
See \cite{res} for an update on linear response around nonequilibria in continuous time.

% -----------------------------------------------

\end{document}